\documentclass{emulateapj}  
  
\shorttitle{Parameters determination using thermal component} \shortauthors{Pe'er et. al.}

\newcommand{\keV}{\rm{\, keV }}

\newcommand{\beq}{\begin{equation}}  
\newcommand{\eeq}{\end{equation}}  
\newcommand{\ba}{\begin{array}}  
\newcommand{\ea}{\end{array}}

\newcommand{\D}{{\mathcal {D}}}  
\newcommand{\R}{{\mathcal {R}}}  
  
\def \etal{{\it et al.~}}

\begin{document}  
\title{A new method of determining the initial size and Lorentz factor  
 of gamma-ray burst fireballs using a thermal emission component}  
  
\author{Asaf Pe'er\altaffilmark{1}, Felix Ryde\altaffilmark{2}, Ralph A.M.J. Wijers\altaffilmark{1}, Peter M\'esz\'aros\altaffilmark{3} and Martin  
 J. Rees\altaffilmark{4} }  
  
\altaffiltext{1}{Astronomical Institute ``Anton Pannekoek'',  
Kruislaan 403, 1098SJ Amsterdam, the Netherlands;  
apeer@science.uva.nl} \altaffiltext{2}{Dept. of Physics, Royal  
Institute of Technology, AlbaNova, SE-106 91 Stockholm, Sweden}  
\altaffiltext{3}{Dept. of Astron. \& Astrophysics, Dept. of  
Physics, Pennsylvania State University, University Park, PA 16802}  
\altaffiltext{4}{Institute of Astronomy, University of Cambridge,  
Madingley Rd., Cambridge CB3 0HA, UK}  
  
\begin{abstract}  
  
In recent years increasing evidence has emerged for a thermal  
component in the $\gamma$- and X-ray spectrum of the prompt  
emission phase in gamma-ray bursts. The temperature and flux of  
the thermal component show a characteristic break in the temporal  
behavior after a few seconds. We show here, that measurements of  
the temperature and flux of the thermal component at early times  
(before the break) allow the determination of the values of two of  
the least restricted fireball model parameters: the size at the  
base of the flow and the outflow bulk Lorentz factor. Relying on  
the thermal emission component only, this measurement is  
insensitive to the inherent uncertainties of previous estimates of  
the bulk motion Lorentz factor. We give specific examples of the  
use of this method: for GRB970828 at redshift $z=0.9578$, we show  
that the physical size at the base of the flow is $r_0 =  
(2.9\pm1.8)\times 10^8\, Y_0^{-3/2}$~cm and the Lorentz factor of  
the flow is $\Gamma = (305\pm28)\, Y_0^{1/4}$, and for GRB990510  
at $z=1.619$, $r_0=(1.7\pm1.7)\times 10^8 \, Y_0^{-3/2}$~cm and  
$\Gamma=(384\pm71) \, Y_0^{1/4}$, where $Y = 1 Y_0$ is the ratio  
between the total fireball energy and the energy emitted in  
$\gamma$-rays.  
\end{abstract}  
  
  
\section{Introduction}  
\label{sec:intro}  
 In recent years increasing evidence has appeared that, during  
the first stages of the prompt emission of long duration gamma-ray  
bursts (GRBs), a thermal component accompanies the underlying  
non-thermal emission [\citet{Ryde04, Ryde05, Camp06}; see also  
\citet{GCG03, KPB03}]. \footnote{It was claimed by \citet{Ru04,  
Br05} that afterglow emission  
  can also be explained with a thermal component.}  
An analysis of BATSE bursts  
that are dominated by quasi-thermal emission \citep{Ryde04, Ryde05}  
showed that the observed temperature exhibits a similar behavior in  
all of them: an initially (approximately) constant temperature  
at a canonical value $T^{ob.}_0 \simeq 100 \keV$ which after $\sim  
1-3$~seconds decreases as a power law in time $T^{ob.} \propto  
t^{-\alpha}$, with power law index $\alpha \simeq 0.6 - 1.1$.  The  
redshifts of most of these bursts are unknown.  An additional analysis  
\citep{RP07} shows that after a short rise, the flux of the black body  
component of these bursts also decreases with time as $F_{BB}^{ob.}  
\propto t^{-\beta}$, with $\beta \simeq 2.0 - 2.5$.  We  
showed there that this temporal behavior can be explained  
as due to the high latitude emission phenomenon \citep{FMN96, GPS99,  
  Qin02}.  
  
According to the standard fireball scenario, the non-thermal  
photons originate from the dissipation of the fireball kinetic  
energy. The dissipation mechanism (e.g., internal shocks  
\citep{PX94, RM94}, magnetic reconnection \citep{GS05,G06} or  
external shocks \citep{MR93,DM99}) is yet uncertain, and can in  
principle occur at various locations. As opposed to this ambiguity  
in understanding the origin of the non-thermal component, the  
thermal component must originate at the photosphere. According to  
the high latitude emission interpretation of the data, the highest  
temperature and the maximal thermal flux initially observed are  
emitted from the photosphere on the radial axis towards the  
observer. Thus, in principle the radius of the emission site of  
these photons can be determined.  
  
In this {\it Letter} we show that combined early time measurements  
of the observed temperature and thermal flux for bursts with known  
redshift allow us to directly determine the values of the bulk  
motion Lorentz factor, the physical size at the base of the flow  
and the photospheric radius.  This is due to the fact that the  
observed temperature and flux of the thermal component depend on  
three internal parameters only: the isotropic equivalent  
luminosity of the thermal component $L_{BB}$, the Lorentz factor  
of the bulk motion of the flow at the photospheric radius $\eta$  
and the physical size at the base of the flow $r_0$, and that  
$L_{BB}$ can be directly measured for bursts with known redshift  
and measured thermal flux.  In \S\ref{sec:model} we give a short  
description of the model, and implications are given in  
\S\ref{sec:implications}. In \S\ref{sec:discussion} we summarize  
and compare our results with those of previous methods of  
estimations of the bulk motion Lorentz factor.  
  
\section{Model: Extended photospheric emission}  
\label{sec:model} In the classical fireball model of gamma-ray  
bursts \citep{Good86,Pac86,PX94}, a thermal plasma of electrons,  
positrons, and photons expands rapidly from an initial radius  
$r_0$. Conservation of energy and entropy imply that the bulk  
Lorentz factor of the flow increases as $\Gamma(r) \propto r$,  
until the plasma reaches the saturation radius $r_s = \eta r_0$,  
above which the plasma Lorentz factor coasts with $\Gamma = \eta  
\equiv L/{\dot M} c^2$.  Here, $L$ is the isotropic equivalent  
burst luminosity, ${\dot M}$ is the mass ejection rate and  
$c$ is the speed of light (from here on we restrict the discussion  
to long bursts, characterized by extended emission of relativistic  
wind).  
  
The photospheric radius $r_{ph}$ is the radius above which the flow
becomes optically thin to scattering by the baryon related
electrons. Depending on the values of the free model parameters
($\eta$, $L$ and $r_0$) this radius can be smaller or larger than
$r_s$ \citep{MRRZ02}: for $\eta > (<) \eta_* \equiv (L \sigma_T/4 \pi
m_p c^3 r_0)^{1/4}$, $r_{ph}<(>) r_s$. Here, $\sigma_T$ is the Thomson
cross section and $m_p$ is the proton mass. The luminosity $L$ is
measured for bursts with known redshift, $L=4\pi d_L^2 Y F^{ob.}$,
where $d_L$ is the luminosity distance, $F^{ob.}$ is the total
(thermal + non thermal) observed $\gamma$-ray flux, and $Y \equiv
\epsilon/\epsilon_\gamma \geq 1$ is the ratio between the total
fireball energy and the energy emitted in $\gamma$-rays. As we show
below, the measurement of $r_0$ is similar in both scenarios,
$r_{ph}>(<) r_s$. Thus, it is possible to determine 
whether $r_{ph}$ is below or above $r_s$ from measurable quantities,
and to determine $\eta$ in the second case.
  
The thermal component originates from the photosphere of an  
expanding plasma jet. The observed thermal flux (integrated over  
all frequencies) is given by integrating the intensity over the  
emitting surface, $F^{ob.}_{BB} = (2\pi/d_L^2) \int d\mu \mu  
r_{ph}^2 \D^4 (\sigma T'^4 / \pi)$. Here, $T'$ is the comoving  
temperature at the photospheric radius, $\sigma$ is Stefan's  
constant and $\D = \D(\theta)$ is the Doppler factor, $\D =  
[\Gamma(1-\beta \mu)]^{-1}$. The angle $\theta$ is the angle  
between the direction of the outflow velocity vector ($\beta$) and  
the line of sight, $\mu \equiv \cos(\theta)$, and $\Gamma = (1-  
\beta^2)^{-1/2}$ is the outflow Lorentz factor.  For a plasma  
Lorentz factor much larger than the inverse of the jet opening  
angle $\Gamma \gg \theta_j^{-1}$ and for early enough times, at a  
given observed time the integration boundaries are determined  
uniquely by the emission duration, regardless of the value of  
$\theta_j$.

Due to the Doppler effect and to the cosmological redshift, photons
that are emitted at frequency $\nu'$ in the comoving frame of a
relativistically expanding plasma with Lorentz factor $\Gamma$ at
redshift $z$, are observed at frequency $\nu^{ob.} = \D (\theta) \nu'
/ (1+z)$. For $\Gamma \gg 1$ and emission on the line of sight,
$\nu^{ob.} \simeq 2 \Gamma \nu'/(1+z)$. In the extended emission
interpretation of the data \citep{RP07}, an observer sees
simultaneously photons that originate from a range of angles to the
line of sight, $\theta_{\min} \leq \theta \leq \theta_{\max}$. In this
case, a thermal spectrum (with temperature $T'$) in the comoving frame
is observed as a modified black-body spectrum. Nonetheless, the
observed spectrum is very similar to a pure black-body spectrum
\citep{Peer07}.

During the first few seconds, when the observed temperature is nearly  
constant, the observed radiation is dominated by photons emitted close  
to the line of sight, i.e., $\theta_{\min} = 0$. According to this  
interpretation, at $t^{ob.}>t_{break}$, there is no more emission from  
$\theta=0$ because the inner engine activity decreases, and the emission  
is dominated  by high latitude effects. Here, $t_{break}$ is the break  
time in the temperature's temporal behavior.  
At $t^{ob.}<t_{break}$, the observed  
spectrum is very close to a black body spectrum with temperature  
$T^{ob.} \simeq 1.48 \Gamma T' / (1+z)$ \citep{Peer07}\footnote{The  
relation between $T^{ob.}$ and $T'$ is often written in the literature  
as $T^{ob.} \simeq \Gamma T'/(1+z)$.  In fact, $\D(\theta=1) \simeq 2  
\Gamma$, thus one should add an extra factor of 2. The factor 1.48  
used here results from the angular integration. See \citet{Peer07} for  
details.}. During this period, the upper integration boundary in the  
equation for the thermal flux is $\mu_{\max} = \cos(\theta_{\min})=1$,  
and the ratio $(F^{ob.}_{BB} / \sigma {T^{ob.}}^4)^{1/2}$ which we  
denote as $\R$, is equal to  
\beq  
\R \equiv \left({F^{ob.}_{BB} \over  
\sigma {T^{ob.}}^4}\right)^{1/2} = (1.06) {(1+z)^2 \over d_L} {r_{ph}  
\over \Gamma}.  
\label{eq:R}  
\eeq  
The prefactor $(1.06)$ originates from the dependence of the  
photospheric radius on the angle to the line of sight \citep{Peer07}.  
  
We can now make the discrimination between the two possible cases:  
$r_{ph} < (>) r_s$. If $r_{ph} < r_s$, then $\Gamma(r) \propto r$. In  
this case, $r_{ph}/\Gamma = r_0$, and equation \ref{eq:R} becomes  
\beq  
r_0 (r_{ph}<r_s) = {1 \over (1.06)} {d_L \over (1+z)^2} \R .  
\label{eq:r0_s}  
\eeq  
In this case, it is not possible to determine the  
photospheric radius, or the value of $\Gamma(r_{ph})$.

In the case $r_{ph}> r_s$, the  
photospheric radius is given by $r_{ph} = (L \sigma_T / 8\pi \eta^3  
m_p c^3)$ \citep{MRRZ02,DM02,Broderick05}.  
At this radius, the comoving temperature is given by  
\beq  
T'(r_{ph}) = \left({L \over 4 \pi r_0^2 c a}\right)^{1/4}  
\eta^{-1} \left({r_{ph} \over r_s}\right)^{-2/3},  
\label{eq:calc_Tw}  
\eeq  
where $a$ is the radiation constant.  
Setting $\Gamma=\eta$ and $L = 4 \pi d_L^2 Y F^{ob.}$ in the equation of  
the photospheric radius, one obtains from equation \ref{eq:R} the coasting  
value of the Lorentz factor,  
\beq  
\eta = \left[(1.06) { (1+z)^2 d_L} {Y F^{ob.} \sigma_T \over 2 m_p  
    c^3 \R} \right]^{1/4}.  
\label{eq:eta}  
\eeq  
Equations \ref{eq:R}, \ref{eq:calc_Tw} and \ref{eq:eta} now give  
the physical size at the base of the flow,  
\beq  
r_0(r_{ph}>r_s) = {4^{3/2} \over (1.48)^6 (1.06)^4} {d_L \over  
  (1+z)^2} \left({F^{ob.}_{BB} \over Y F^{ob.}}\right)^{3/2} \R.  
\label{eq:r0_g}  
\eeq  
  
We thus find that a measurement of $\R$ and the ratio of the black  
body flux to the total flux at the very early observed times from  
bursts with known redshift give a direct measurement of $r_0$, and  
that the result is similar (up to a numerical factor of the order  
unity, provided that $Y$ is not much larger than 1; see discussion  
on the value of $Y$ in \S\ref{sec:discussion}), for the two  
considered cases, $r_{ph}< (>) r_s$. The measured values of $r_0$  
and $L$ can be used to determine the value of $\eta_*$, which is  
independent of the specific scenario. One can then use the  
measured values of $\R$ and $F^{ob.}$ to determine the value of  
$\eta$ using equation \ref{eq:eta}.  If the obtained value is  
larger than the value of $\eta_*$, then $r_{ph}< r_s$, in which  
case equation \ref{eq:eta} should not be used and the value of  
$\eta$ remains undetermined.

\section{Implications}  
\label{sec:implications}  
  
Relations \ref{eq:eta} and \ref{eq:r0_g} allow a direct measurement of  
the size at the base of the flow, $r_0$ and of the bulk motion Lorentz  
factor of the flow from GRBs with known redshift and energy content.  
In addition, it is  
possible to determine the photospheric radius $r_{ph}$ and the  
saturation radius $r_s$, if measurements of the thermal flux and  
temperature are available at early enough times.  
  
We illustrate the use of this method on two bursts with known  
redshifts, namely GRB970828 and GRB990510 observed by the BATSE  
detectors aboard the {\it Compton Gamma Ray Observatory}. BATSE  
detected bursts in the 20 keV -- 2 MeV energy range. The  
time-resolved spectra for the selected bursts were fitted using a  
Planck function and a single power-law to model the photospheric and  
the non-thermal emission components, following the method presented  
in \citet{Ryde04} \citep[see also][]{RP07}.  
  
\begin{figure}  
\plotone{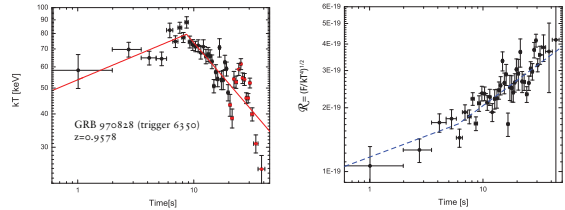}  
 \caption{Temporal behavior of the thermal component in GRB970828 at  
$z=0.9578$. Left panel: the observed temperature. During the first  
$\sim 8$~s the temperature rises slowly up to $78.5 \keV$, after  
which it decreases as a power law in time with power law index  
$-0.51$. The fit was made on the data up to 20 s, which includes  
the first pulse structure of the light curve. Right panel: the  
temporal behavior of the ratio $\R \equiv (F_{BB}/\sigma  
{T^{ob}}^4)^{1/2}$. This ratio increases slowly  during the first  
$\sim 7$~s, after which it increases as a power law in time with  
power law index $0.67$. The break time after $\sim 7$~s is close  
to the break time in the temporal behavior of the temperature. }  
\label{fig1}  
\end{figure}

The analysis of the thermal component of GRB970828 is presented in
figures \ref{fig1}. This burst, at redshift $z=0.9578$, had a good
temporal coverage for the first 100 seconds. The left hand panel in
Figure \ref{fig1} shows the temporal behavior of the temperature of
the thermal component. During the first $\sim 8$~seconds the observed
temperature rises slightly to a value of $78.5 \keV$, after which it
shows a rapid decrease that can be fitted as a power law in time with
a power law index $\alpha = -0.51$. In the right hand panel of Figure
\ref{fig1} we show the temporal behavior of the function $\R$. This
function shows a slight increase during the first $\sim 7$~seconds,
after which it rises as a power law in time with a power law index
$\beta = 0.67$. The smooth increase of $\R$ before the break implies
that there is no significant energy dissipation below the photosphere,
which, if it occurred would result in a strong fluctuation and affect
the smoothness of $\R$.  The break times in the temporal behavior of
the observed temperature and $\R$ are the same within the errors.
  
According to the extended high latitude emission interpretation of
this result, we deduce that the first episode of significant inner
engine activity took place during the first $7-8$ seconds, and at
later times we are observing photons emitted off axis (we neglect here
late time episodes of engine activity that occur after $\sim 25$~s and
$\sim 60$~s in this burst). The values of $r_0$ and $\eta$ are
calculated using the observed values of the temperature $T^{ob.} =
78.5 \pm 4.0$~\keV, $\R= (1.88 \pm 0.28) \times 10^{-19}$, and the
ratio of thermal to total flux $F^{ob.}_{BB}/F^{ob.} = 0.64\pm0.20$ at
the break time.  The error bars on the measured quantities are
averaged over the first seconds, before the temporal break.
Considering a flat universe with $\Omega_\Lambda = 0.73$, $H_0 = 71
{\rm \, km/s/Mpc}$, the luminosity distance for this burst is $d_L =
1.94\times 10^{28}$~cm.  Using equations \ref{eq:eta} and
\ref{eq:r0_g}, we find that $\Gamma = (305\pm28) \, Y_0^{1/4}$ and
$r_0 =(2.9\pm1.8) \times 10^8 \, Y_0^{-3/2}$~cm, where $Y = 1
Y_0$. The calculated value of $\eta_* = 463 \, Y_0^{5/8}$ proves that
indeed $r_{ph} = 2.7\times 10^{11} \, Y_0^{1/4}$~cm is larger than
$r_s=9.0\times 10^{10} \, Y_0^{-5/4}$~cm, which implies that $\eta =
\Gamma = 305 \, Y_0^{1/4}$ is the coasting value of the outflow
Lorentz factor. The statistical error on the estimated value of $\eta$
is $\lesssim 10\%$. The systematical error results from the
uncertainty in the value of $Y$, and is not expected to be more than
tens of percents (see \S\ref{sec:discussion} below), giving the best
constraint on the estimated value of $\eta$ measured so far.
  
A similar analysis was carried out for GRB990510 at $z=1.619$. The
results obtained are similar, $r_0=(1.7 \pm 1.7)\times 10^8 \,
Y_0^{-3/2}$~cm and $\eta=(384\pm71) \, Y_0^{1/4}$. The larger
statistical errors compared to GRB970828 mainly reflect the fewer
available data points for this burst. The value of $\eta_* = 830\,
Y_0^{5/8}$ proves that indeed $r_{ph}=7.7\times 10^{11} \,
Y_0^{1/4}$~cm is larger than $r_s = 6.3\times 10^{10} \, Y_0^{-5/4}$.
The temporal behavior of the observed temperature and $\R$ were found
to be similar in a large sample of BATSE bursts, providing further
evidence for our model. However, the redshift of most of these bursts
is unknown, thus definite values of $\eta$ and $r_0$ could not be
obtained. The full sample appears in \citet{RP07}.

\section{Discussion}  
\label{sec:discussion}  
  
In this {\it Letter} we showed that by measuring the observed  
temperature and thermal flux of the thermal component that  
accompanies the prompt emission of GRBs, it is possible to  
determine the values of two of the least restricted parameters of  
the fireball model: the size at the base of the flow and the  
outflow bulk Lorentz factor (in the case that the photospheric  
radius is larger than the saturation radius). In this case, it is  
also possible to determine the saturation radius $r_s$ and the  
photospheric radius $r_{ph}$.  We showed that the calculation of  
the initial size of the flow (equations \ref{eq:r0_s},  
\ref{eq:r0_g}) is similar (up to a constant of the order unity)  
for the cases $r_{ph} > (<) r_s$. This allows a comparison between  
the measured value of $\eta$ (equation \ref{eq:eta}) and the  
derived value of $\eta_*$, and a discrimination between the two  
cases. We have given examples of the use of this method for the  
determination of $\eta$ and $r_0$ for two specific GRBs.  
  
The largest uncertainty in the estimate of $\eta$ is due to the
uncertainty in the value $Y$. Value of $1 \lesssim Y < 3 - 5$
were suggested based on afterglow observations \citep{FW01,Frail01,
FB05, GKP06}. Theoretical arguments based on fitting the flux of
ultra-high energy cosmic rays \citep[UHECR; ][]{WDA04}, under the
assumption that UHECRs originate from GRBs \citep{W95, Vietri95}, as
well as efficiency considerations \citep{FP06} suggest larger value,
$Y\gtrsim10$.  For bursts with a dominant thermal component, as
considered here, we expect $Y$ to be close to unity. If afterglow
measurements of bursts with detected thermal component and known
redshift become available, as expected after the launch of the {\it
GLAST} satellite, this uncertainty could be removed. A second source
of systematic uncertainty can result from dominating Compton
scattering (which conserves the number of photons) resulting in a Wien
spectrum, rather than a thermal spectrum.  Observationally, a Wien
spectrum is hard to discriminate from a thermal spectrum. In this
case, the systematic error in estimating the temperature is $(3 k_B T)
/ (2.7 k_B T) \sim 10\%$, which transforms into $\sim 5\%$ uncertainty
in the estimated value of $\eta$. Currently, this uncertainty is
smaller than the statistical uncertainty, $\gtrsim 10\%$.

Other methods of estimating the bulk motion Lorentz factor in GRBs  
relied on a large number of uncertain model assumptions and  
uncertainties in the values of the free model parameters. A widely  
used lower limit for $\eta$ is obtained by calculating the minimum  
Lorentz factor required in order for the observed energetic photons  
not to annihilate \citep{KP91, FEH93, WL95, BH97, LS01}. In addition  
to providing only a lower limit, in order to get a good estimate of  
$\eta_{\min}$ a wide spectral coverage of the GRB emission, from the  
optical band to the $\gamma$-rays is required.
In these calculations, the prompt emission spectrum  is sometimes
approximated as a broken power law \citep[e.g.,][]{LS01}, which may be
too simplified \citep[e.g.,][]{PW04}.

An alternative method to estimate the Lorentz factor is by  
modeling the early afterglow emission, on the assumption that the  
optical flash observed in a few cases results from synchrotron  
emission by electrons heated by the reverse shock \citep{SP99}. A  
serious drawback of this method is that the estimate relies on the  
poorly known shock microphysics parameters (such as $\epsilon_e$,  
$\epsilon_B$ etc.). A more advanced method, introduced by  
\citet{ZKM03}, relies on comparing the emission from the forward  
and reverse shock during the early afterglow. An underlying  
assumption in this estimate is that the values of the microphysics  
parameters at the forward and the reverse shocks are similar.  
Other methods rely on measurements of the physical parameters  
during the late afterglow emission, assuming that the flow expands  
in a self-similar motion during this phase. The initial value of  
the Lorentz factor is deduced by measuring the rise time of the  
early afterglow \citep{S99, WDL00, SR02, KZ03}. An inherent  
drawback of this method is the assumption that the microphysical  
parameters are constant in time during the late afterglow.  
  
The method presented here of estimating $\eta$ is independent of  
any of the uncertainties inherent in the former methods. Moreover,  
it gives a direct measurement of $\eta$, rather than a lower  
limit. The results presented in \S\ref{sec:implications}  
\citep[see also][]{RP07} indicate values of $\eta$ close to the  
earlier estimates.  In addition, the values found for the size at  
the base of the flow $r_0$ could further constrain GRB progenitor  
models. The statistical errors on the values of these numbers are  
much smaller than any previous estimates.  
  
These facts have several important consequences.  First, they
strengthen the interpretation of the prompt emission as being composed
of a thermal component, in addition to the non-thermal
component. Therefore, any interpretation of the prompt emission data
must take this thermal component into account. Second, it shows that
the extended high latitude emission interpretation of the late time
temporal behavior of the thermal component is consistent with the
fireball model predictions. Thus, this interpretation may also be used
to understand the strong X-ray flares observed by the {\it SWIFT}
satellite. Third, the consistency found between the different methods
for estimating the value of $\eta$ can be used to strengthen the
validity of the underlying assumptions in previous estimates of
$\eta$, such that the values of the microphysical parameters
($\epsilon_e$, $\epsilon_B$, etc.) are indeed constant in time during
the afterglow emission phase. 
And last, the direct measurement of the physical size at the
base of the flow is another, independent indication that a massive
star is indeed the progenitor of long duration GRBs. Since $\eta$ is
related to the mass ejection rate, our measurements could be useful to
constrain models of GRB progenitors.

\acknowledgements This research was supported by NWO grant  
639.043.302 to R.W.. F.R. acknowledges the support by the Swedish  
National Space Board. P.M. wishes to acknowledge the support by  
NASA NAG5-13286 and NSF AST0307376. This research made use of data  
obtained through the HEASARC Online Service provided by the NASA  
Goddard Space Flight Center. F.R. wishes to express his gratitude  
to the Department of Astronomy at Amsterdam University for their  
hospitality. We wish to thank the anonymous referee for useful  
comments.

\end{document}